\documentclass[%
 reprint,
 amsmath,amssymb,
 aps,
 prl,
]{revtex4-2}

\usepackage{graphicx}
\usepackage{dcolumn}
\usepackage{bm}
\usepackage{subcaption}
\usepackage{placeins}  
\usepackage{verbatim}
\usepackage{amsmath}
\usepackage{amssymb}
\usepackage{algpseudocode}
\usepackage{algorithm}
\usepackage{physics}
\usepackage{xcolor}
\usepackage[colorlinks=true, citecolor=teal, linkcolor=red, urlcolor=magenta]{hyperref}
\usepackage{cleveref}

\floatstyle{ruled}
\restylefloat{algorithm}
\newcommand{\parhead}[1]{\textit{#1.---}}

\makeatletter
\long\def\@makecaption#1#2{%
  \par
  \vskip\abovecaptionskip
  {\small\rmfamily
   \leftskip\z@ \rightskip\z@ \parfillskip\@flushglue \parindent\z@
   \def\@caption@fignum@sep{: }%
   \@make@capt@title{#1}{#2}\par}%
  \vskip\belowcaptionskip}
\makeatother

\begin{document}
\raggedbottom

\preprint{APS/123-QED}

\title{Unbiased sampling from Boltzmann distributions with noisy energies}

\author{Iwo Sanderski}
\author{Gian Gentinetta}
\author{Giuseppe Carleo}
\affiliation{Institute of Physics, \'Ecole Polytechnique F\'ed\'erale de Lausanne (EPFL), CH-1015 Lausanne, Switzerland}
\affiliation{Center for Quantum Science and Engineering, EPFL, Lausanne, Switzerland}

\date{\today}

\begin{abstract}
Sampling from the Boltzmann distribution is central to computational physics, yet hard when the energy is known only through a stochastic estimate, such as with machine-learned molecular potentials, in variational Monte Carlo, or on quantum computers, because a noisy energy biases the sampled distribution. The penalty method of Ceperley and Dewing~\cite{ceperley_penalty_1999} corrects this but requires the noise variance and becomes intractable when it is large. We introduce the Poisson product estimator, an unbiased, non-negative estimator of the Boltzmann weight that only needs an upper bound on the energy estimator and remains efficient at high noise. Using it to optimize a variational quantum circuit gradient-free, we recover the $H_3^+$ ground-state energy in a minimal basis and, by sampling rather than following a single trajectory, also map the variational energy landscape.

\end{abstract}

\maketitle

Sampling from the Boltzmann distribution of a many-body system is a central task in computational physics, underpinning the evaluation of thermodynamic and equilibrium properties~\cite{mermin_thermal_1965, georges_dynamical_1996, white_finite-temperature_2020}. Because a low-temperature distribution concentrates on the lowest-energy configurations, the same sampling tools also offer a route to optimization, for instance through simulated annealing~\cite{kirkpatrick_optimization_1983, cerny_thermodynamical_1985}, or, in variational ground-state search, by drawing low-energy parameters of an ansatz~\cite{ritz_uber_1909, peruzzo_variational_2013, carleo_solving_2017} rather than relying on gradient descent.

In many settings of interest, however, the energy is not available exactly but only through a stochastic estimator. This situation is widespread across computational physics, for instance in classical and semiclassical molecular simulations driven by machine-learned or coarse-grained potentials that carry a predictive uncertainty~\cite{jinnouchi_phase_2019}, or by energies obtained from an inner stochastic calculation; in variational Monte Carlo (VMC), where the energy is a statistical average over sampled configurations~\cite{mcmillan_ground_1965, becca_quantum_2017}; and on quantum computers, where the energy of a many-body Hamiltonian is exponentially expensive to obtain exactly and is instead estimated from a finite number of measurements~\cite{peruzzo_variational_2013}. While Markov chain Monte Carlo (MCMC) techniques such as the Metropolis-Hastings algorithm enable sampling from Boltzmann distributions without knowing the partition sum~\cite{metropolis_equation_1953, hastings_monte_1970}, they assume access to the energy at a given configuration. Replacing the exact energy in the acceptance ratio with an (unbiased) energy estimator results in a biased Boltzmann distribution in general~\cite{ceperley_penalty_1999}.

Ceperley and Dewing~\cite{ceperley_penalty_1999} remove this bias by modifying the acceptance probability with a \textit{penalty} term that depends on the variance of energy estimator. This variance has to be either known or estimated stochastically leading to an additional sampling overhead. They show that while the method is efficient for small noise variances, the acceptance rate is exponentially suppressed with the variance, making the simulation infeasible in regimes that naturally arise in VMC and quantum-computing applications.

\begin{figure*}
    \includegraphics[width=\linewidth]{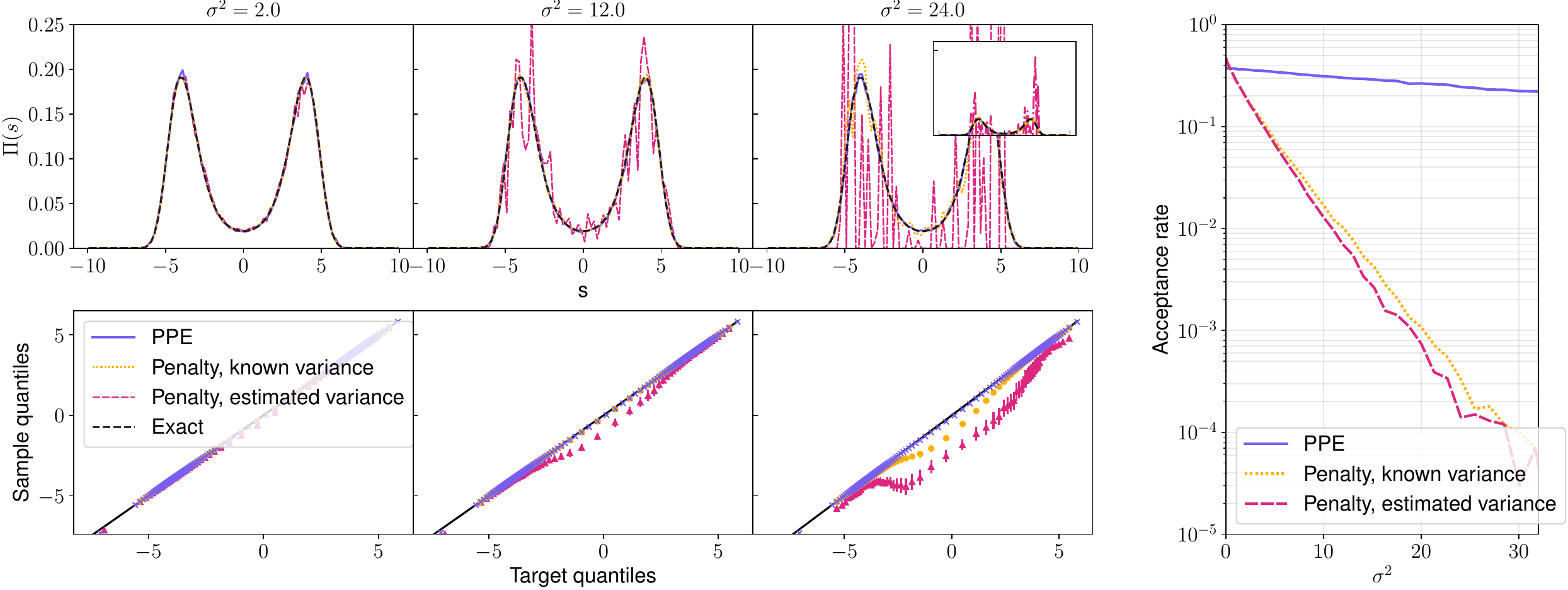}
    \caption{Sampling the Boltzmann distribution of the double-well potential with noisy energies. We compare the PPE (purple) with the penalty method using the exact noise variance (``Penalty, known variance''; yellow) and its variant that estimates the variance stochastically through the Bessel correction (``Penalty, estimated variance''; red). \textit{Top row:} histograms of the sampled distributions overlaid on the exact Boltzmann density, for energy-estimate variances $\sigma^2 = 2.0, 12.0, 24.0$ (left to right; inset zooms the $\sigma^2=24.0$ peaks). \textit{Bottom row:} the corresponding quantile-quantile plots against the exact distribution. While at $\sigma^2 = 2.0$ all three estimators are accurate, the penalty methods deviate strongly at larger variances, whereas the PPE reproduces the exact distribution throughout. \textit{Right:} the acceptance rate versus noise variance collapses exponentially for the penalty methods but stays high for the PPE.}
    \label{fig:double_well_samples}
\end{figure*}

In this work, we propose an alternative approach: an unbiased, non-negative estimator of the Boltzmann weight based on a Poisson-product representation of the exponential. This construction is closely related to the Poisson estimator used for unbiased Monte Carlo integration~\cite{beskos_exact_2006, wagner_unbiased_1987} and to pseudo-marginal and Russian-roulette Markov chain Monte Carlo~\cite{andrieu_pseudo-marginal_2009, lyne_russian_2015}. Unlike the penalty method, which corrects the bias only under a Gaussian-noise assumption and requires knowledge or estimation of the noise variance, our estimator is unbiased for any noise distribution and needs neither. Its sole requirement is an almost-sure upper bound on the energy estimator, which keeps all product factors non-negative. Such a bound cannot be dispensed with in general~\cite{jacob_nonnegative_2015}, but it is automatically and exactly satisfied by the shot-based estimators used on quantum computers, making the method naturally suited to that setting.

In this letter, we apply the estimator to two complementary settings. First, we show that it outperforms the penalty method in the double-well potential from \cite{ceperley_penalty_1999} when the noise is large. We then combine it with simulated annealing of a variational quantum circuit to approximate the ground state of the $H_3^{+}$ ion. We recover the full-configuration-interaction (FCI) ground-state energy in this minimal basis to high precision and with high probability, which shows that our approach offers a gradient-free alternative to the variational quantum eigensolver (VQE) \cite{peruzzo_variational_2013}.

Our aim is to sample states $\{s_i\}_{i=1,\dots,M}$ from the distribution $s_i \sim e^{-\beta E(s_i)}$, where the energy value $E(s)$ can only be estimated stochastically through an unbiased estimator
\begin{equation}
 E(s) = \mathbb{E}[\epsilon(s)] \approx \frac{1}{K}\sum_{k=1}^K \epsilon_k(s).
\end{equation}

The nonlinearity of the exponential function introduces a bias when using this estimator to sample from the Boltzmann distribution in general \cite{ceperley_penalty_1999, ceperley_review}. To circumvent that problem, one can either add a correction to the acceptance rate as proposed by~\cite{ceperley_penalty_1999}, or directly construct an unbiased estimator of the Boltzmann distribution.

\parhead{Overview of the penalty method}The penalty method devised by Ceperley and Dewing~\cite{ceperley_penalty_1999} adds a \textit{penalty} term  
\begin{equation}
    u(\sigma^2) = \beta^2\sigma^2,
    \label{eq:direct_penalty}
\end{equation} 
depending on the variance $\sigma^2:=\text{Var}[\epsilon]$ to the computation of the acceptance ratio in the Metropolis-Hastings algorithm. The resulting expression
\begin{equation}
    a_P(s \to s'; \sigma) = \min\left\{1,\ \exp\!\left(-\beta\bigl(\epsilon(s') - \epsilon(s)\bigr)-u(\sigma^2)\right)\right\}
\label{eq:penalty_acceptance}
\end{equation}
compensates for the bias due to noisy energy estimates. 
When the variance of the noise $\sigma^2$ is not known exactly, the penalty term is replaced by the \textit{Bessel penalty}
\begin{equation}
    u_B(\chi^2) = \beta^2\chi^2 + \frac{\beta^4\chi^4}{n+1} + \dots,
    \label{eq:bessel_penalty}
\end{equation}
where the variance $\chi^2$ is now estimated stochastically.

Both penalty terms will exponentially suppress the acceptance rate, rendering the method highly inefficient at large noise levels.
\begin{figure*}
    \includegraphics[width = \linewidth]{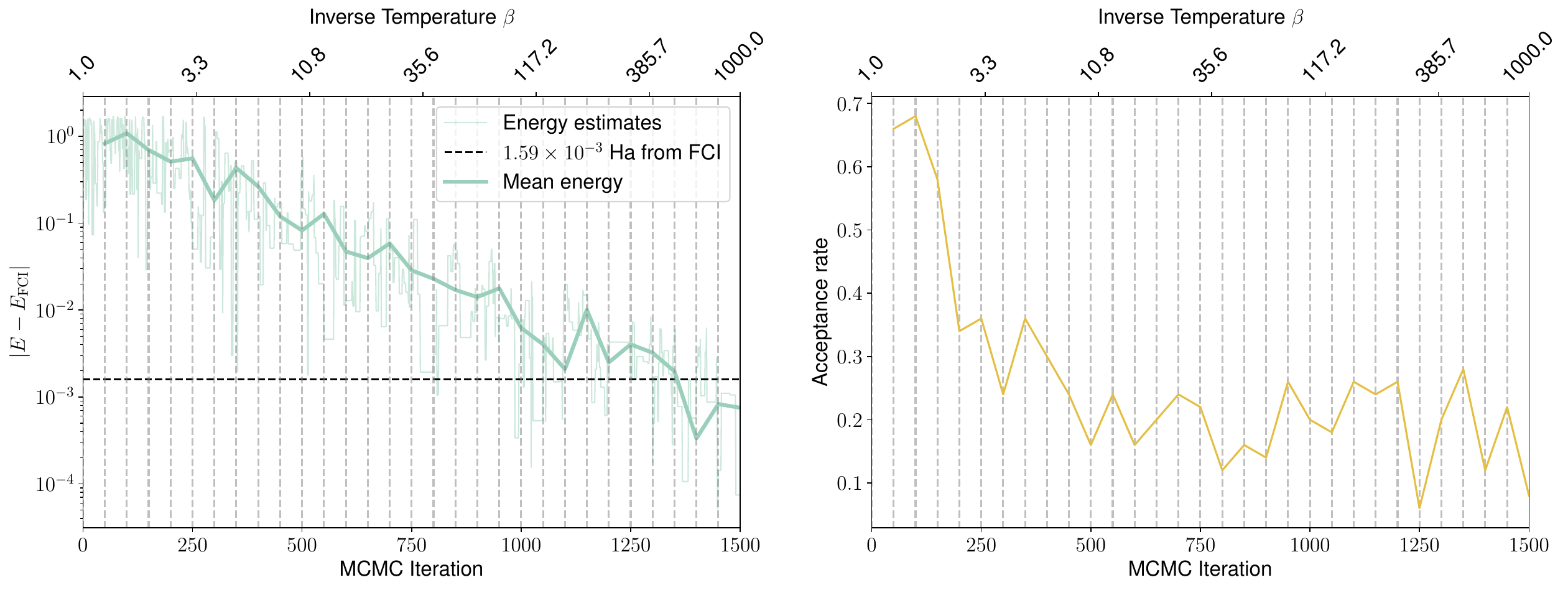}
    \caption{Optimization of a variational quantum circuit through simulated annealing with the PPE. We approach the ground state of the $H_3^+$ molecule in a minimal basis set by sampling variational parameters $\theta$ from the Boltzmann distribution $e^{-\beta E(\theta)}$ for an increasing inverse temperature $\beta$. We plot the energy estimates $\epsilon(\theta_t)$ as a function of the Markov chain Monte Carlo iteration $t$ as well as the mean energy reached for each value of $\beta$. As the system cools down, the energy decreases, converging to the in-basis FCI ground-state energy around $\beta=1000$. On the right, we plot the mean acceptance rate for each value of $\beta$ of the MCMC simulation.}
    \label{fig:energy_and_acceptance}
\end{figure*}

\parhead{The Poisson product estimator}Rather than correcting the acceptance ratio, we construct an unbiased, non-negative estimator of the Boltzmann weight $e^{-\beta E(s)}$ itself. The construction rests on one elementary identity: if $N$ is drawn from a Poisson distribution $P_\lambda$ of rate $\lambda$, the average of $x^N$ is exponential in $x$,
\begin{equation}
    \mathbb{E}_{N\sim P_\lambda}\!\left[x^{N}\right] = \sum_{n=0}^{\infty} e^{-\lambda}\frac{\lambda^{n}}{n!}\,x^{n} = e^{\lambda(x-1)} .
    \label{eq:poisson_pgf}
\end{equation}
Choosing $x = 1 - \beta(E(s)-C)/\lambda$ makes the right-hand side equal to $e^{\beta C}e^{-\beta E(s)}$, i.e.\ the Boltzmann weight up to a constant. 

The variable $x$ is linear in the energy, so although $E(s)$ is unknown it can be estimated by the noisy samples. We note that averaging over the single factors
\begin{equation}
    g(\epsilon) = 1 - \frac{\beta(\epsilon - C)}{\lambda}
\end{equation}
results in the mean $\mathbb{E}[g(\epsilon)] = x$, since $\epsilon$ is an unbiased estimate of $E(s)$. We therefore draw $N \sim P_\lambda$ together with $N$ independent energy estimates $\{\epsilon_k(s)\}_{k=1}^{N}$, and multiply the factors to form the \emph{Poisson product estimator}
\begin{equation}
    \hat{\pi}(s) = \prod_{k=1}^{N} g(\epsilon_k(s)) .
    \label{eq:ppe}
\end{equation}
Because the $\epsilon_k$ are independent, $\mathbb{E}[\hat\pi(s)\,|\,N] = x^{N}$. Averaging over $N$ with \cref{eq:poisson_pgf} gives
\begin{equation}
    \mathbb{E}[\hat{\pi}(s)] = e^{\beta C}\,e^{-\beta E(s)} ,
    \label{eq:ppe_unbiased}
\end{equation}
as derived in full in the End Matter. The prefactor $e^{\beta C}$ is independent of $s$ and cancels in the Metropolis acceptance ratio, so $\hat\pi$ is an unbiased stand-in for the Boltzmann weight and can be sampled with a pseudo-marginal Metropolis chain~\cite{andrieu_pseudo-marginal_2009} (Algorithm~\ref{alg:poisson_pseudocode}, End Matter).

For $\hat\pi$ to be a valid sampling weight it must be non-negative~\cite{jacob_nonnegative_2015}. Each factor satisfies $g(\epsilon_k)\geq 0$ only when $\beta(\epsilon_k-C)\leq\lambda$; guaranteeing this for every sample requires an upper bound $\epsilon(s)\leq E_{\text{max}}$ together with a rate
\begin{equation}
    \lambda \geq \beta(E_{\text{max}} - C) .
    \label{eq:lambda_lower_bound}
\end{equation}
For unbounded noise this hard bound is replaced in practice with one that holds with high probability only, leaving a controllable tail bias. The shift $C$ is otherwise free. Placing it near the typical energies keeps each factor close to unity and minimizes the variance (End Matter).

\parhead{Ground states via Boltzmann sampling}Our algorithm can be combined with a simulated annealing schedule~\cite{ kirkpatrick_optimization_1983, cerny_thermodynamical_1985} to approximate the ground state of quantum many-body systems such as molecules. Given a variational ansatz $\ket{\psi(\theta)}$ and Hamiltonian $H$, parameters $\theta$ corresponding to energies $E(\theta) = \bra{\psi(\theta)}H\ket{\psi(\theta)}$ close to the ground-state energy can be obtained with high probability by sampling $\theta \sim e^{-\beta E(\theta)}$
for sufficiently large $\beta$. 

Directly sampling a low-temperature distribution with Algorithm~\ref{alg:poisson_pseudocode} becomes increasingly difficult with the dimension of the parameter space. This is because the distribution is highly peaked at low temperatures, rendering the chance of choosing the initial parameters in a low-energy regime vanishingly small~\cite{Arrasmith_2022}.

To avoid this issue we start sampling from a high-temperature distribution at $\beta_0$. We then progressively increase the inverse temperature,  $\beta_{i} < \beta_{i+1}$, until the target $\beta_{n_{\beta}}$ is reached (Algorithm~\ref{alg:annealing_pseudocode}). This procedure  contracts the target distribution towards the global minimum of the variational energy and avoids having to locate the narrow low-temperature peak directly.

Finally, the ground state energy is estimated from the last $k$ sampled parameters that describe wave functions close to the ground state. The energy is obtained by averaging the energy estimates $\epsilon(\theta_k)$, and, following the variational principle, we keep the lowest energy obtained.

This is particularly interesting when $\ket{\psi(\theta)}$ is a variational quantum circuit, where gradient-based optimization is often costly due to expensive gradient estimation, or exponentially vanishing gradients ~\cite{mcclean_barren_2018, abbas_quantum_2023, holmes_connecting_2021}.
The simulated annealing approach has the advantage of not requiring gradient estimation. 
\begin{figure*}
    \includegraphics[width=\textwidth]{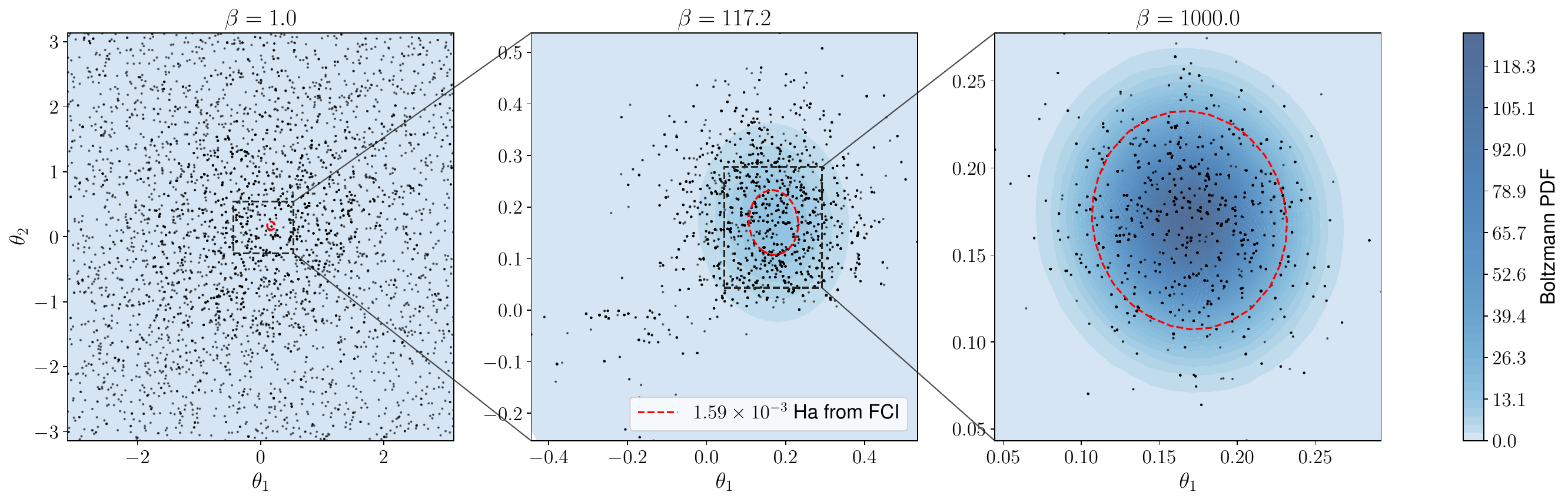}
    \caption{Parameter-space points sampled from the Boltzmann distribution with the PPE coupled to an annealing schedule, for three values of the inverse temperature $\beta$. The panels at $\beta = 117.2$ and $\beta = 1000$ are zoomed relative to the one at $\beta = 1$. The red circle marks the region in which the variational energy lies within $1.59\times10^{-3}$~Ha of the FCI minimum. As the system cools, the Boltzmann distribution concentrates in the low-energy region, so that at $\beta=1000$ the sampled parameters fall within this tolerance of the minimum with high probability.}
    \label{fig:sampling_boltzmann_visualized}
\end{figure*}

\parhead{Double-well potential}We compare the penalty method to the PPE using the Boltzmann distribution at $\beta = 1.0$ of the double-well potential presented in \cite{ceperley_penalty_1999}
\begin{equation}
    V(s) = -0.288s^2 + 0.009s^4.
\end{equation}
We add artificial Gaussian noise of variance $\sigma^2$ to each evaluation of the energy function $E(s)$
\begin{equation}
    \epsilon(s) = E(s) + \mathcal{N}(0, \sigma^2),
\end{equation}
which provides an unbiased estimator of the energy that can be used in Algorithm \ref{alg:poisson_pseudocode}. This Gaussian model is a controlled stand-in for the stochastic energy estimates met in practice, for instance from a machine-learned potential with predictive uncertainty or from an inner stochastic calculation, and lets us vary the noise level systematically.

We benchmark the PPE against both variants of the penalty method---using the exact noise variance, and the Bessel correction when the variance is unknown---at a fixed budget of $2\times10^{7}$ energy evaluations per method, updating configurations by a Gaussian random-walk proposal (per-method sample counts and parameters in the Supplemental Material).

\Cref{fig:double_well_samples} compares the sampled distributions to the exact Boltzmann density at three noise levels, $\sigma^2 \in \{2.0, 12.0, 24.0\}$. At low noise all three estimators are accurate, but as the variance grows both penalty variants develop a pronounced bias, evident as systematic deviations in the quantile-quantile plots, while the PPE stays faithful to the target throughout. The Bessel variant, which does not presume the variance to be known and is therefore the relevant comparison in practice, degrades the fastest.

This contrast originates in the acceptance rate (right panel of \Cref{fig:double_well_samples}). The penalty enters the acceptance probability as $e^{-\beta^2\sigma^2}$ [\Cref{eq:penalty_acceptance}] and is thus exponentially suppressed in the variance, falling below $10^{-2}$ and starving the chain of accepted moves. The PPE acceptance also decreases, but only slowly. The relative variance of its weight grows as $e^{\beta^2\sigma^2/\lambda}$ (End Matter), a growth that the Poisson rate $\lambda$ holds in check. The estimator therefore keeps sampling accurately in a regime where the penalty method has become impractical.

\parhead{Molecular ground states}We now use the annealed PPE to find the ground state of the $H_3^+$ ion with a variational quantum circuit $\ket{\psi(\theta)} = U(\theta)\ket{0}$, annealing along a geometric schedule $\beta_{i+1} = \eta\beta_i$. The molecular Hamiltonian is mapped to qubits through the Jordan-Wigner transformation~\cite{jordan_uber_1928},
\begin{equation}
\label{eq:pauli_decomp}
    H = \sum_{i} \alpha_i P_i, \qquad \alpha_i \in \mathbb{R}.
\end{equation}

The energy $\langle H\rangle_{\psi(\theta)}$ is estimated from measurements of the circuit. Rather than measuring all $\mathcal{O}(n^4)$ Pauli terms of an $n$-orbital Hamiltonian, we draw one term per shot with probability $|\alpha_i|/\gamma$, where $\gamma = \sum_i|\alpha_i|$, giving the unbiased estimator
\begin{align}
    E(\theta) &= \langle H\rangle_{\psi(\theta)} = \mathbb{E}_{i\sim |\alpha_i|/\gamma}[\gamma \cdot\text{sgn}(\alpha_i)\cdot \langle P_i\rangle_{\psi(\theta)}]\\
    &= \mathbb{E}_{i\sim |\alpha_i|/\gamma} \left[\mathbb{E}_{s \sim p(s|\theta, P_i)}[\epsilon_{s, P_i}] \right],
\end{align}
with $s \in \{-1, 1\}$ the measured eigenvalue of $P_i$ and $\epsilon_{s, P_i} = \gamma \cdot\text{sgn}(\alpha_i)\cdot s$. Being a single Pauli outcome, this estimator is bounded, $|\epsilon_{s, P_i}| \leq \gamma$, so $E_{\text{max}} = \gamma$ is an exact bound. The PPE hence samples without bias and without any truncation, in contrast to the unbounded Gaussian noise above. Its variance is reduced by averaging $b$ shots per estimate,
\begin{equation}
    \epsilon(\theta) = \frac{1}{b}\sum_{k=1}^b \gamma\, \text{sgn}(\alpha_{i_k})\,s_k, \quad s_k\sim p(s|\theta, P_{i_k}), \quad i_k \sim \frac{|\alpha_i|}{\gamma},
    \label{eq:vqe_batched_estimators}
\end{equation}
which lowers the weight fluctuations and raises the acceptance rate (End Matter). The full simulation setup is given in the Supplemental Material.

\Cref{fig:energy_and_acceptance} shows the annealing run for the circuit of Ref.~\cite{Utkarsh2023Chemistry}. As $\beta$ increases the sampled energies descend toward the FCI ground state, converging near $\beta = 1000$, while the mean acceptance rate decreases but remains finite throughout. Evaluating the lowest-energy parameters from the cold end of the chain exactly gives a variational energy of $-1.26203$~Ha, within $1.4\times10^{-4}$~Ha of the STO-3G FCI value; the sampler thus locates the exact ground state of this minimal-basis Hamiltonian. Beyond locating the minimum, sampling returns the entire parameter distribution. This thermodynamically weighted map of the variational energy landscape is not provided traditionally through gradient-based and other point-estimate optimizers that follow a single trajectory. Because this circuit has only two parameters, that map can be visualized directly. \Cref{fig:sampling_boltzmann_visualized} shows the parameters drawn by the annealed PPE at three inverse temperatures, against the exact Boltzmann distribution from state-vector simulation (here using $5000$ samples per temperature for clarity). The sampled and exact landscapes agree closely even at large $\beta$, and as the system cools the distribution contracts onto the low-energy region, until at $\beta = 1000$ the samples lie within $1.59\times10^{-3}$~Ha of the FCI minimum (red circle). The PPE produces this map gradient-free and despite the stochastic energies. The full simulation setup is given in the Supplemental Material.

\parhead{Conclusion}The PPE is an unbiased, non-negative estimator of the Boltzmann distribution when the energy is accessible only through stochastic estimation. Sampling from it with the Metropolis-Hastings algorithm achieves far higher acceptance than the penalty method of Ref.~\cite{ceperley_penalty_1999} once the energy variance is large. Unlike that method, the PPE is unbiased for arbitrary bounded noise and needs no estimate of the variance. This is especially relevant for the bounded, non-Gaussian estimators native to quantum hardware.

Combined with simulated annealing, the estimator minimizes a variational energy in a gradient-free manner. For a quantum circuit it recovers the FCI ground state of $H_3^+$ in a minimal basis by cooling to $\beta = 1000$, with the energy estimated stochastically from sampled Pauli terms rather than the full Hamiltonian expectation value. Moreover, by drawing the entire parameter distribution rather than following a single trajectory, it maps the variational energy landscape, and a potential probe of cost-landscape features such as barren plateaus and narrow gorges~\cite{mcclean_barren_2018, Arrasmith_2022}. A detailed comparison to gradient-based and gradient-free VQE optimizers, such as the simultaneous perturbation stochastic approximation~\cite{spall_multivariate_1992, kandala_hardware-efficient_2017}, in a competitive, large-scale setting remains an open direction.

More broadly, sampling from a Boltzmann distribution whose energy is known only stochastically is a recurring problem across computational physics, and the two applications shown here are only specific instances of it. We expect the PPE to be useful wherever the energy is noisy but bounded, such as in classical and semiclassical molecular simulations driven by machine-learned or coarse-grained potentials with predictive uncertainty~\cite{jinnouchi_phase_2019}, in variational Monte Carlo, and in quantum-computing applications. In all of these, it offers an unbiased, variance-agnostic alternative to the penalty method, and we envision it as a subroutine for Metropolis sampling with uncertain energies.

\section*{Acknowledgements}
We thank Markus Holzmann for fruitful discussions. This research was supported by the NCCR MARVEL, a National Centre of Competence in Research, funded by the Swiss National Science Foundation (grant number 205602).

\bibliography{notes}

\appendix
\begin{widetext}
\begin{center}
\rule{0.5\textwidth}{0.4pt}\\[0.4em]
{\large\textbf{End Matter}}\\[0.2em]
\rule{0.5\textwidth}{0.4pt}
\end{center}
\medskip
\section{Algorithms}
\label{appdx:pseudocodes}
\begin{algorithm}[H]
\caption{Poisson product estimator sampler}
\label{alg:poisson_pseudocode}
\begin{algorithmic}[1]
\Require Hamiltonian operator H, Inverse temperature $\beta$, Upper bound $ E_{\text{max}}$ of the energy estimate, Constant shift $C$, Number of desired samples $M$, Initial state $s\sim U(\Omega)$ where $\Omega$ is the state space.
\Ensure $S$: $M$ samples from the Boltzmann distribution $\pi(s) = \exp(-\beta E(s))$
\vspace{2mm}
\State Initialize $S \gets[\,]$
\State Define $\lambda \gets \beta(E_{\text{max}} - C)$
\State Draw $N \sim P_{\lambda}(N)$
\State Draw $\epsilon_1(s),\dots, \epsilon_N(s) \gets \Call{GetEstimator}{H, s, N}$  
\State Compute $\hat{\pi}(s) \gets \prod_{k=1}^N \left( 1-\frac{\beta(\epsilon_k(s) -C)}{\lambda}\right)$
\For{$n = 1$ \textbf{to} $M$}
    \State Draw proposal $s' \sim T(s \to s')$ 
    \State Draw $N' \sim P_{\lambda}(N)$
    \State Draw $\epsilon_1(s'),\dots, \epsilon_{N^{'}}(s') \gets \Call{GetEstimator}{H, s', N'}$
    \State Compute $\hat{\pi}(s') \gets \prod_{k=1}^{N'} \left( 1-\frac{\beta(\epsilon_k(s') -C)}{\lambda}\right)$
    \State Compute ratio $r \gets \dfrac{\hat{\pi}(s')\,T(s' \to s)}{\hat{\pi}(s)\,T(s \to s')}$
    \State Set acceptance $\alpha \gets \min(1, r)$
    \State Draw $u \sim U([0,1])$
    \If{$u < \alpha$}
        \State $s \gets s'$ \Comment{Accept}
        \State $\hat{\pi}(s) \gets \hat{\pi}(s')$ 
    \Else
        \State $s \gets s$ \Comment{Reject}
    \EndIf
    
    \State \Call{Append}{$S$, $s$}
\EndFor
\State Return $S$
\end{algorithmic}
\end{algorithm}

\begin{algorithm}[H]
\caption{Simulated annealing}
\label{alg:annealing_pseudocode}
\begin{algorithmic}[1]
\Require Every input required by Algorithm~\ref{alg:poisson_pseudocode} ($H$, $C$, $E_{\text{max}}$, $M$, initial state $s$), besides $\beta$; a list of inverse temperatures $\boldsymbol{\beta}$ of length $n_{\beta}$.
\Ensure $S$: $M/n_{\beta}$ samples from each Boltzmann distribution $\pi(s) = \exp(-\beta_i E(s))$.
\vspace{2mm}
\State Initialize $S \gets[\,]$
\For{$n = 1$ \textbf{to} $n_{\beta}$}
\State $S_n \gets \Call{PPE-Sampler}{H, \beta_n, C, E_{\text{max}}, M/n_{\beta}, s}$ \Comment{Algorithm~\ref{alg:poisson_pseudocode}}
\State $s \gets S_n[\text{end}]$
\State \Call{Append}{$S$, $S_n$}
\EndFor
\State Return $S$
\end{algorithmic}
\end{algorithm}
\section{Proof of the unbiasedness of the Poisson product estimator}
\label{appdx:unbiasedness_estimator}
The energy estimates $\epsilon(s)$ for a configuration $s$ are drawn from a probability density $q_s(\epsilon)$ with mean $E(s) = \mathbb{E}_{\epsilon \sim q_s}[\epsilon]$. Since $N \sim P_{\lambda}(N)$ and the $\epsilon_k$ are drawn independently from $q_s$, the expectation of the estimator factorizes. Introducing
\begin{equation}
  m_{s} \;:=\; \mathbb{E}_{\epsilon \sim q_s}\!\left[g(\epsilon)\right] \;=\; 1-\frac{\beta\left(E(s)-C\right)}{\lambda},
\end{equation}
the expectation conditioned on $N$ reads $\mathbb{E}[\hat{\pi}(s,N)\,|\,N] = \left(\mathbb{E}_{\epsilon \sim q_s}[g(\epsilon)]\right)^{N} = m_{s}^{N}$. Averaging over the Poisson distribution of $N$, and using the series for $e^{z}$ together with $\lambda(1-m_{s}) = \beta\left(E(s)-C\right)$, then yields
\begin{equation}
  \mathbb{E}_{N}[\hat{\pi}(s, N)] \;=\; \sum_{n=0}^{\infty} m_{s}^{\,n}\, e^{-\lambda}\,\frac{\lambda^{n}}{n!} \;=\; e^{-\lambda(1-m_{s})} \;=\; \exp\!\bigl[-\beta\left(E(s)-C\right)\bigr] \;=\; e^{\beta C}\,e^{-\beta E(s)} .
\end{equation}
Since the factor $e^{\beta C}$ is independent of $s$, it cancels in the Metropolis acceptance ratio, so $\hat{\pi}$ is an unbiased estimator of the Boltzmann weight.

\section{Variance and sampling efficiency}
\label{appdx:variance}
The efficiency of pseudo-marginal sampling is governed by the fluctuations of the weight $\hat\pi$, which should be kept of order unity~\cite{andrieu_pseudo-marginal_2009}. The moments of the estimator can be obtained in closed form. Writing $a \equiv \beta/\lambda$ and $g(\epsilon) = 1 - a(\epsilon - C)$, with $\epsilon \sim q_s$ of mean $E \equiv E(s)$ and variance $\sigma^2$, the first two moments of $g$ are
\begin{align}
    m_1 &= \mathbb{E}_q[g] = 1 - a(E-C), \\
    m_2 &= \mathbb{E}_q[g^2] = 1 - 2a(E-C) + a^2\bigl[(E-C)^2 + \sigma^2\bigr].
\end{align}
Since $N \sim P_\lambda$ and the $\epsilon_k$ are independent, $\mathbb{E}[\hat\pi^{\,p}] = \exp\!\bigl[\lambda(\mathbb{E}_q[g^p]-1)\bigr]$. For $p=1$ this reproduces the unbiasedness result $\mathbb{E}[\hat\pi] = e^{-\beta(E-C)}$, while $p=2$ yields the exact relative variance
\begin{equation}
    \frac{\mathrm{Var}[\hat\pi]}{\mathbb{E}[\hat\pi]^2} = \exp\!\left(\frac{\beta^2\bigl[(E-C)^2 + \sigma^2\bigr]}{\lambda}\right) - 1 .
    \label{eq:relative_variance}
\end{equation}
To leading order, the exponent equals $\mathrm{Var}[\log\hat\pi]$. \Cref{eq:relative_variance} clarifies the role of every hyperparameter. The shift $C$ should track the energy to minimize $(E-C)^2$, the variance $\sigma^2$ can be lowered by batching [\Cref{eq:vqe_batched_estimators}], and increasing the rate $\lambda$ directly suppresses the fluctuations. Keeping the exponent of order unity requires $\lambda = \mathcal{O}\!\bigl(\beta^2[(E-C)^2 + \sigma^2]\bigr)$, i.e. a number of estimator evaluations per Metropolis step that grows polynomially with the inverse temperature.

\newpage
\begin{center}
\rule{0.5\textwidth}{0.4pt}\\[0.4em]
{\large\textbf{Supplementary material}}\\[0.2em]
\rule{0.5\textwidth}{0.4pt}
\end{center}
\medskip
\section{Double-well benchmark}
\label{appdx:double_well}
For the double-well benchmark of the main text we fix a budget of $2\times10^7$ energy evaluations per method. The penalty method spends two evaluations per Metropolis step (current and proposed configuration), giving $n_{s,\text{pen}} = 1\times10^7$ samples; the Bessel variant additionally uses $n = 64$ evaluations per step to estimate the variance, giving $n_{s,b} = 156250$ samples. For the PPE the noisy energies lie with high probability in $[E_{\text{min}}, E_{\text{max}}] = [-5, 25]$; choosing $C = E_{\text{min}}$, each step draws on average $\lambda = E_{\text{max}} - C = 30$ evaluations of the proposed configuration, giving $n_{s,p} = 6.7\times10^5$ samples. Configurations are updated by a Gaussian random-walk proposal $s' = s + \mathcal{N}(0, \sigma_{\text{prop}}^2)$ with $\sigma_{\text{prop}} = 3.0$.

\section{Computing setup to simulate the \texorpdfstring{$H_3^+$}{H3+} ion and performance improvements}
\label{appdx:molecule_setup}
In this section we provide the details of the simulation of the $H_3^+$ molecule reported in the main text.
The Hamiltonian is defined on 6 qubits and is created by mapping the second-quantized fermionic Hamiltonian describing the molecule in the STO-3G minimal basis set to qubits using the Jordan-Wigner transformation~\cite{jordan_uber_1928} through PennyLane~\cite{arrazola2021differentiable}. The ansatz, taken from~\cite{Utkarsh2023Chemistry}, consists of two Double Excitation gates, applied on qubits \{1, 2, 3, 4\} and \{1, 2, 5, 6\}, applied to the initial Hartree-Fock state.

As described in the main text, we run a geometric annealing schedule where $\eta = 1.268$ is chosen such that the values of $\beta$ increase from $\beta_1 = 1.0$ to $\beta_{n_{\beta}} = 1000.0$ in 30 steps. At each inverse temperature we sample $N_s = 50$ points using the Poisson product estimator, proposing new points by a Gaussian random walk, $\theta' = \theta + \mathcal{N}(0,\sigma_{\text{prop}}^2)$, independently for each coordinate with
\begin{equation}
    \sigma_{\text{prop}, i} = 0.1\cdot\sqrt{\frac{\beta_{n_\beta}}{\beta_i}}.
\end{equation}
We observed that scaling the proposal width as $1/\sqrt{\beta}$ achieves better performance compared to a constant $\sigma_{\text{prop}}$.

We found that the hyper-parameter with the highest impact on overall performance is the constant shift $C$. This constant should be chosen such that it is close to the energy estimations $\epsilon_k(s)$ in order to achieve the best performance. For small absolute differences $| \epsilon_k(s) - C|$, the factors $g(\epsilon_k(s))$ are close to unity, which leads to better stability and higher acceptance rates. In practice, it is not always possible to choose an ideal shift $C$. For instance, the estimator $\epsilon_{s, P_i} = \gamma \cdot\text{sgn}(\alpha_i)\cdot s$ used to estimate the energy of the Pauli Hamiltonian can only take values of $\pm \gamma$. In this case, there is no shift $C \in \mathbb{R}$ such that $|C - \epsilon_k(s)|$ is small for all samples. To remedy this issue, one can average the estimator over batches, as defined in the main text, to reduce the variance. In our simulation of the $H_3^+$ molecule we apply this technique with batch size $b = 1000$, which combined with choosing the constant shift as the Hartree-Fock energy $C = -1.237$ Ha results in sufficient acceptance rates.

Since the batched estimator is bounded by $|\epsilon(\theta)| \leq \gamma$, we use the exact upper bound $E_{\text{max}} = \gamma$, which guarantees non-negativity of the PPE without any truncation. Finally, we observed that it can be beneficial to choose $\lambda > \beta (E_{\text{max}} - C)$. This further increases the acceptance rate at the cost of having more intermediate samples. For the simulations of the $H_3^+$ molecule we decided to pick $\lambda = 2\cdot \beta(E_{\text{max}} - C)$.

\end{widetext}

\end{document}